\documentclass[runningheads]{llncs}
\usepackage{subcaption}
\usepackage{graphicx}
\newcommand{\vect}[1]{\boldsymbol{#1}}
\usepackage{subfiles}
\usepackage{wrapfig}
\usepackage{hyperref}
\usepackage{amsmath,amssymb}
\DeclareMathOperator{\E}{\mathbb{E}}
\usepackage{float}
\graphicspath{ {./images/} }
\usepackage[utf8]{inputenc}
\usepackage[english]{babel}

\begin{document}

\title{Alzheimer's Disease Modelling and Staging through Independent Gaussian Process Analysis of Spatio-Temporal Brain Changes}
\titlerunning{Independent Gaussian Process Analysis}

\author{Clement Abi Nader\inst{1} \and Nicholas Ayache\inst{1} \and Philippe Robert\inst{2,3} \and Marco Lorenzi\inst{1}, for the Alzheimer's Disease Neuroimaging Initiative$^{*}$}

\institute{UCA, Inria Sophia Antipolis, Epione Research Project \and UCA, CoBTeK \and Centre Memoire, CHU de Nice}

\authorrunning{C. Abi Nader et al.}

\maketitle  
\begin{abstract}Alzheimer's disease (AD) is characterized by complex and largely unknown progression dynamics affecting the brain's morphology. Although the disease evolution spans decades, to date we cannot rely on long-term data to model the pathological progression, since most of the available measures are on a short-term scale. It is therefore difficult to understand and quantify the temporal progression patterns affecting the brain regions across the AD evolution. In this work, we tackle this problem by presenting a generative model based on probabilistic matrix factorization across temporal and spatial sources. The proposed method addresses the problem of disease progression modelling by introducing clinically-inspired statistical priors. To promote smoothness in time and model plausible pathological evolutions, the temporal sources are defined as monotonic and independent Gaussian Processes. We also estimate an individual time-shift parameter for each patient to automatically position him/her along the sources time-axis. To encode the spatial continuity of the brain sub-structures, the spatial sources are modeled as Gaussian random fields. We test our algorithm on grey matter maps extracted from brain structural images. The experiments highlight differential temporal progression patterns mapping brain regions key to the AD pathology, and reveal a disease-specific time scale associated with the decline of volumetric biomarkers across clinical stages.

\end{abstract}

\let\thefootnote\relax\footnote{*Data used in preparation of this article were obtained from the Alzheimer’s Disease Neuroimaging Initiative (ADNI) database (adni.loni.usc.edu). As such, the investigators within the ADNI contributed to the design and implementation of ADNI and/or provided data but did not participate in analysis or writing of this report. A complete listing of ADNI investigators can be found at: http://adni.loni.usc.edu/wp-content/uploads/how\textunderscore to\textunderscore apply/ADNI\textunderscore Acknowledgement\textunderscore List.pdf.}

\section{Introduction}

Neurodegenerative disorders such as Alzheimer's disease (AD) are characterized by morphological and molecular changes of the brain, and ultimately lead to cognitive and behavioral decline \cite{ref_AD}. To date there is no clear understanding of the dynamics regulating the disease progression. Consequently several attempts have been made to model the disease evolution in a data-driven way, using sets of biomarkers extracted from different imaging acquisition techniques, such as Magnetic Resonance Imaging (MRI) \cite{ref_marco}. However available data are mostly represented by cross-sectional measures or time-series acquired on a short-term time span, while the ultimate goal is to unveil the ``long-term'' disease evolution spreading over decades. Therefore there is a critical need to define the AD evolution in a data-driven manner with respect to an absolute time scale associated to the natural history of the pathology. 
\newline
\newline
To this end, in \cite{ref_jedynak} the authors introduce a disease progression score for each patient in order to identify a data-driven disease scale. This score is based on a set of biomarkers and was shown to correlate with the decline of brain cognitive abilities. A similar approach was proposed by \cite{ref_marco} and \cite{ref_donohue} with scalar biomarkers. In \cite{ref_bilgel}, a disease progression score was estimated using higher-dimensional biomarkers from molecular imaging. However these methods don't provide information about the brain structures involved in AD, and how the disease affects them along time. To overcome these limitations, \cite{ref_marinescu} proposes a spatio-temporal model of disease progression explicitly accounting for different temporal dynamics across the brain. This is done by decomposing cortical thickness measurements as a mixture of spatio-temporal processes, by associating each vertex to a temporal progression modeled by a sigmoid function. The approach also estimates a disease progression score for each subject as a linear transformation of time. However since the proposed formulation does not account for spatial correlation between vertices, it may be potentially sensitive to spatial variation and noise, thus leading to poor interpretability. 
\newline
\newline
The challenge of spatio-temporal modelling in brain images is a classical problem widely addressed via Independent Component Analysis (ICA \cite{ref_ica}), especially on functional MRI (fMRI) data \cite{ref_fmri}. ICA aims at decomposing the data via matrix factorization, looking for a reduced number of spatio-temporal latent sources. Although successful in fMRI analysis, ICA cannot find straightforward applications to the modelling of AD progression. First, ICA retrieves maximally independent latent sources best explaining the data. However, although brain regions can exhibit different atrophy rates, this doesn't necessarily imply statistical independence between them. Second, differently from fMRI data, the absolute time axis of AD spatio-temporal observations is unknown. Thus estimating the pathology timing is a key step in order to model the disease progression, and cannot be performed with standard dimensionality reduction methods such as ICA. Finally, fMRI time series are defined over hundreds of time points, while we work essentially in a cross-sectional setting with one or a few images per-subject. 
\newline
\newline
In this work we present a novel spatio-temporal generative model of disease progression aimed at quantifying the independent dynamics of changes in the brain. We model the observed data through matrix factorization across temporal and spatial sources, with a plausibility constraint introduced by clinically-inspired statistical priors. To promote smoothness in time and model steady evolution from normal to pathological stages, the temporal sources are defined as monotonic independent Gaussian Processes (GPs). We also estimate an individual time-shift parameter for each patient to automatically position him along the sources time-axis. To encode the spatial continuity of the brain sub-structures, the spatial sources are modeled as Gaussian random fields. The framework is efficiently optimized through stochastic variational inference. In the next sections we detail the method formulation and show its application on synthetic and real data composed by a large dataset of MRIs from the Alzheimer's Disease Neuroimaging Initiative (ADNI). Further information can be found in the Appendix.

\section{Method}

We assume that the spatio-temporal data $\vect{Y}(x,t) = [\vect{Y_{1}}(x,t_{1}), \vect{Y_{2}}(x,t_{2}),.., \vect{Y_{P}}(x,t_{p})]$ is stored in a matrix with dimensions $P \times F$, where $P$ is the number of patients, $F$ the number of image features, and $\vect{Y_{i}}(x, t_{i})$ is the image of an individual $i$ observed at position $x$ and at time $t_{i}$. We postulate a generative model in order to decompose the data in $N_s$ spatio-temporal sources such that : 
\begin{equation} \displaystyle \vect{Y_{p}}(x,t_{p}) = \vect{S}(\theta, t+t_{p})\vect{A}(\psi, x) + \vect{\mathcal{E}}. \end{equation}
$\vect{S}$ is a $P \times N_s$ matrix where each column represents a temporal trajectory, $t_{p}$ the individual time-shift parameter, and $\theta$ the set of parameters related to the temporal sources. $\vect{A}$ is a $N_s \times F$ matrix where each row represents a spatial map, and $\psi$ is a set of spatial parameters. $\vect{\mathcal{E}}$ is a  $\mathcal{N}(0, \sigma^{2}\vect{I})$ Gaussian noise. According to the generative model the likelihood is : 
\begin{equation}\label{eq:to_optim}p(\vect{Y}|\vect{A},\vect{S}, \sigma) = \displaystyle \prod_{p=1}^{P} \frac{1}{(2\pi \sigma^{2})^\frac{F}{2}} \exp(-\frac{1}{2\sigma^{2}}||\vect{Y_{p}} -\vect{S}(\theta, t+t_{p}) \vect{A}(\psi, x)||^{2}).\end{equation}
\noindent
For each latent source $n$, the row $\vect{A_{n}}$ of $\vect{A}$ is provided with a $\mathcal{N}(0, \vect{I})$ prior, while each column $\vect{S_{n}}$ of $\vect{S}$ is a GP modeled as in \cite{ref_gp}. This setting leverages on kernel approximation through sampling of basis functions in the spectral domain \cite{rahimi}. For specific choices of the covariance, such as the Radial Basis Function (RBF) used in our work, the GPs can be approximated as a Bayesian neural network with form : $\vect{S_{n}}(t) = \phi(\vect{\Omega_{n}} t)\vect{W_{n}}$. According to \cite{ref_gp}, in this formulation $\vect{\Omega_{n}}$ is the projection in the spectral domain provided with a $\mathcal{N}(0, \frac{1}{l_{n}}\vect{I})$ prior, $\phi$ is defined as the $(\cos, \sin)$ function in order to obtain a RBF kernel for the covariance matrix, and the regression parameter $\vect{W_{n}}$ is provided with a $\mathcal{N}(0, \vect{I})$ prior. The GPs are estimated using variational inference by introducing approximated distributions of $\vect{\Omega} = \{ \vect{\Omega_{n}}, n \in [1, N_{s}] \}$ and $\vect{W} = \{ \vect{W_{n}}, n \in [1, N_{s}] \}$. 
\newline
\newline
To account for the steady increase of the sources from normal to pathological stages we introduce a monotonicity prior over the GPs. To do so, we constrain the space of the temporal sources to the set $\mathcal{C} = \{\vect{S}(t)  \mid \vect{S'}(t) \leq 0 \quad \forall t\}$, following \cite{ref_monotonicity}. This leads to a second likelihood term constraining the dynamics of the temporal sources :
\begin{equation} p(\mathcal{C}|\vect{S'}, \lambda) = (1 + \exp(-\lambda \vect{S'}(t)))^{-1}.\end{equation}
\noindent
We jointly optimize \eqref{eq:to_optim} according to priors and constraints, by maximizing the data evidence :
\begin{align} 
\begin{split}
\label{eq:likelihood}
\log(p(\vect{Y}, \mathcal{C}|\sigma, \lambda)) & = \log[\int_{\vect{A}}\int_{\vect{S}} \int_{\vect{S'}}p(\vect{Y}|\vect{A},\vect{S}, \sigma)p(\mathcal{C}|\vect{S'}, \lambda)p(\vect{A})p(\vect{S},\vect{S'}| \lambda)d\vect{A}d\vect{S}d\vect{S'}] \\
& = \log[\int_{\vect{A}}\int_{\vect{S}} \int_{\vect{S'}}p(Y|\vect{A},\vect{S}, \sigma)p(\mathcal{C}|\vect{S'}, \lambda)p(\vect{A})p(\vect{S'}|\vect{S}, \lambda)p(\vect{S})d\vect{A}d\vect{S}d\vect{S'}]. \\
\end{split}
\end{align}
\noindent
By observing that $\vect{S'}$ is completely identified by $\vect{S}$, formula \eqref{eq:likelihood} can be written as :
\begin{align} 
\begin{split}
\label{eq:marginal}
\log(p(\vect{Y},\mathcal{C}|\sigma, \lambda)) & = \log[\int_{\vect{A}}\int_{\vect{S}}p(\vect{Y}|\vect{A},\vect{S}, \sigma)p(\mathcal{C}|\vect{S'}, \lambda)p(\vect{A})p(\vect{S})d\vect{A}d\vect{S}]. \\
\end{split}
\end{align}
Since this integral is intractable, we tackle the optimization of \eqref{eq:marginal} via stochastic variational inference. Following \cite{ref_kingma} and \cite{ref_gp} we introduce approximations $q_1(\vect{A})$, $q_{2}(\vect{\Omega})$ and $q_{3}(\vect{W})$ to derive the lower bound :
\begin{align}
\begin{split}
\label{eq:lower_bound}
\log(p(\vect{Y},\mathcal{C}|\sigma, \lambda)) & \geqslant \E_{\vect{A} \sim q_{1}, \vect{\Omega} \sim q_{2}, \vect{W} \sim q_{3}}[log(p(\vect{Y}|\vect{A},\vect{\Omega}, \vect{W}, \sigma))] + \E_{\vect{\Omega} \sim q_{2}, \vect{W} \sim q_{3}}[log(p(\mathcal{C}|\vect{\Omega},\vect{W}, \lambda))] \\ 
& - \mathcal{D}[q_{1}(\vect{A})||p(\vect{A})] - \mathcal{D}[q_{2}(\vect{\Omega})||p(\vect{\Omega})] - \mathcal{D}[q_{3}(\vect{W})||p(\vect{W})]. 
\end{split}
\end{align}
Where $\mathcal{D}$ refers to the Kullback-Leibler divergence. 
\newline
\newline
We specify the approximated distribution of the spatial activation maps $q_{1}$ such that $q_{1}(\vect{A}) = \textstyle \prod_{n=1}^{Ns} \mathcal{N}(\vect{\mu_{n}}, \vect{\Sigma}(\alpha, \beta))$. To introduce spatial correlations and model a smooth signal decay in the maps across voxels with coordinates $(\vect{u_{i}}, \vect{u_{j}})$, we choose $ \vect{\Sigma}_{i,j}(\alpha, \beta) = \alpha\exp(-||\vect{u_{i}}-\vect{u_{j}}||^{2}/2\beta)$. Moreover, we can use the separability properties of the exponential to decompose the covariance between two locations $\vect{u_{i}} = (x_{i}, y_{i}, z_{i})$ and $\vect{u_{j}} = (x_{j}, y_{j}, z_{j})$ :
\begin{equation}
\vect{\Sigma}_{i,j}(\alpha, \beta) = \alpha\exp(-\frac{(x_{i}-x_{j})^{2}}{2\beta})\exp(-\frac{(y_{i}-y_{j})^{2}}{2\beta})\exp(-\frac{(z_{i}-z_{j})^{2}}{2\beta}). \end{equation}
Thanks to this property $\vect{\Sigma}$ can be decomposed into the Kronecker product of 1D processes, $\vect{\Sigma} = \vect{\Sigma_{x}} \otimes \vect{\Sigma_{y}} \otimes \vect{\Sigma_{z}}$, as shown in \cite{ref_kronecker}, allowing us to deal with large-size matrices.
\newline
\newline
The approximated distributions $q_{2}(\vect{\Omega})$ and $q_{3}(\vect{W})$ are factorized across GPs such that $q_{2}(\vect{\Omega}) = \prod_{n=1}^{N_{s}}q_{2}(\vect{\Omega_{n}}) = \prod_{n=1}^{N_{s}}\prod_{j=1}^{N_{rf}} \mathcal{N}(r_{n,j}, p_{n,j}^{2})$ and $q_{3}(\vect{W}) = \prod_{n=1}^{N_{s}}q_{3}(\vect{W_{n}}) = \prod_{n=1}^{N_{s}}\prod_{j=1}^{N_{rf}} \mathcal{N}(m_{n,j}, s_{n,j}^{2})$, where $N_{rf}$ is the number of random features used for the projection in the spectral domain.
\newline
\newline
Since we only work with Gaussian distributions we can obtain a closed-form for the Kullback-Leibler terms in \eqref{eq:lower_bound}. The expectations are approximated by using the reparameterization trick as presented in \cite{ref_kingma}, and the lower bound is efficiently optimized through backpropagation. We chose to alternate the optimization between the spatio-temporal parameters and the time-shift. We set $\lambda$ to the minimum value that gives monotonic sources, while $\sigma$ was arbitrarily determined from the data. A detailed derivation of the model and lower-bound can be found in the Appendix.

\section{Results}

\subsection{Benchmark on Synthetic Data}
\label{sec:synthetic_data}
\begin{figure}
  \begin{subfigure}[c]{0.44\textwidth}
    \includegraphics[width=\textwidth]{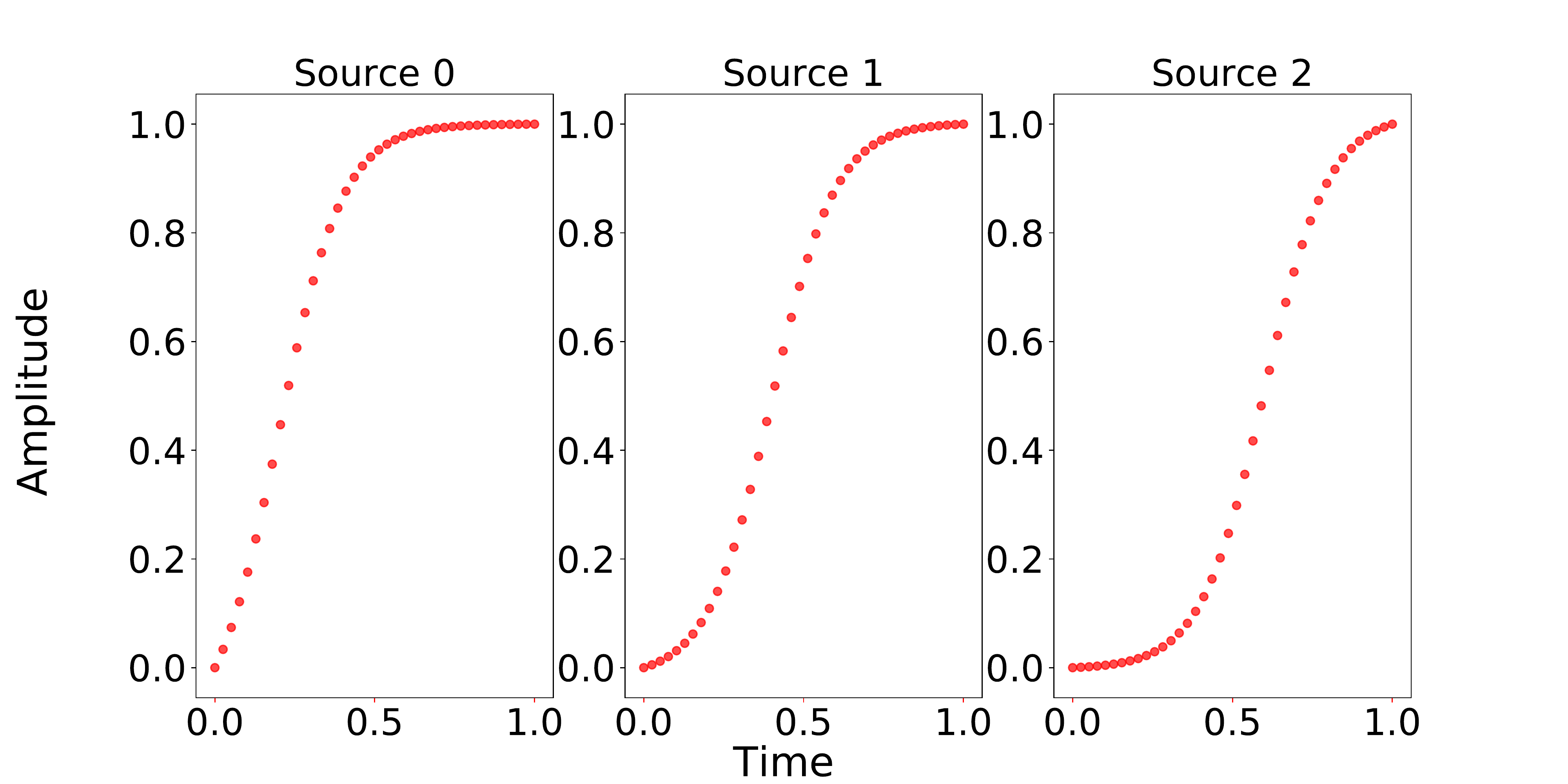}
    \caption{}
    \label{fig:synthetic_S}
  \end{subfigure}
  \begin{subfigure}[c]{0.55\textwidth}
    \includegraphics[width=\textwidth]{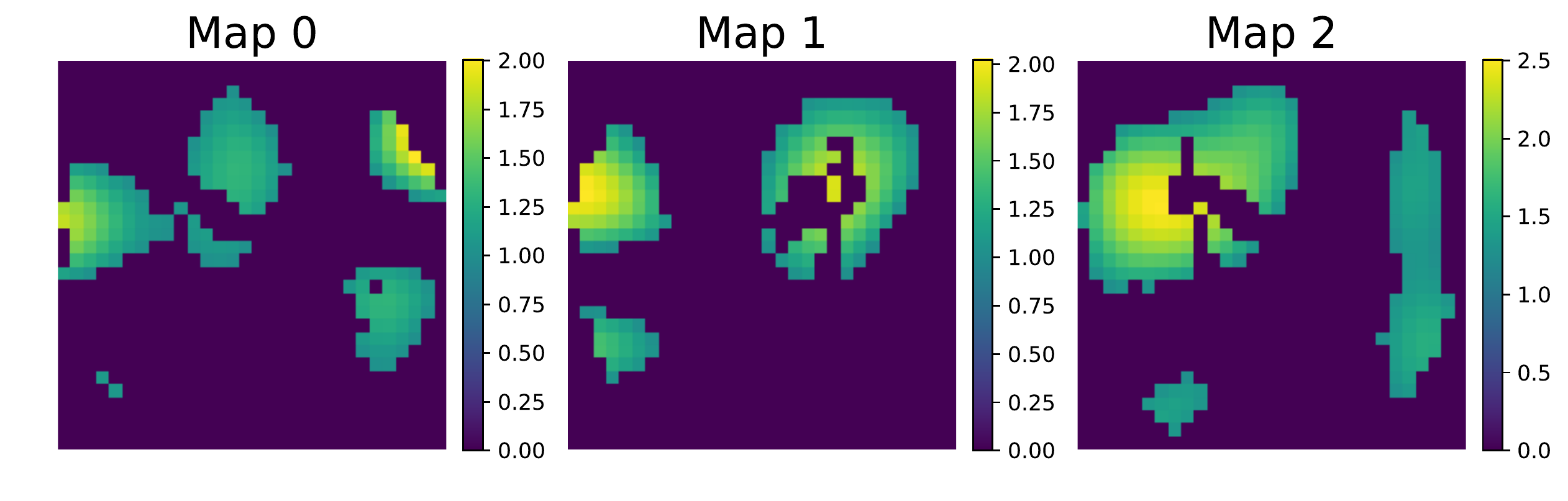}
    \caption{}
    \label{fig:synthetic_A}
  \end{subfigure}
    \begin{subfigure}[c]{0.44\textwidth}
    \includegraphics[width=\textwidth]{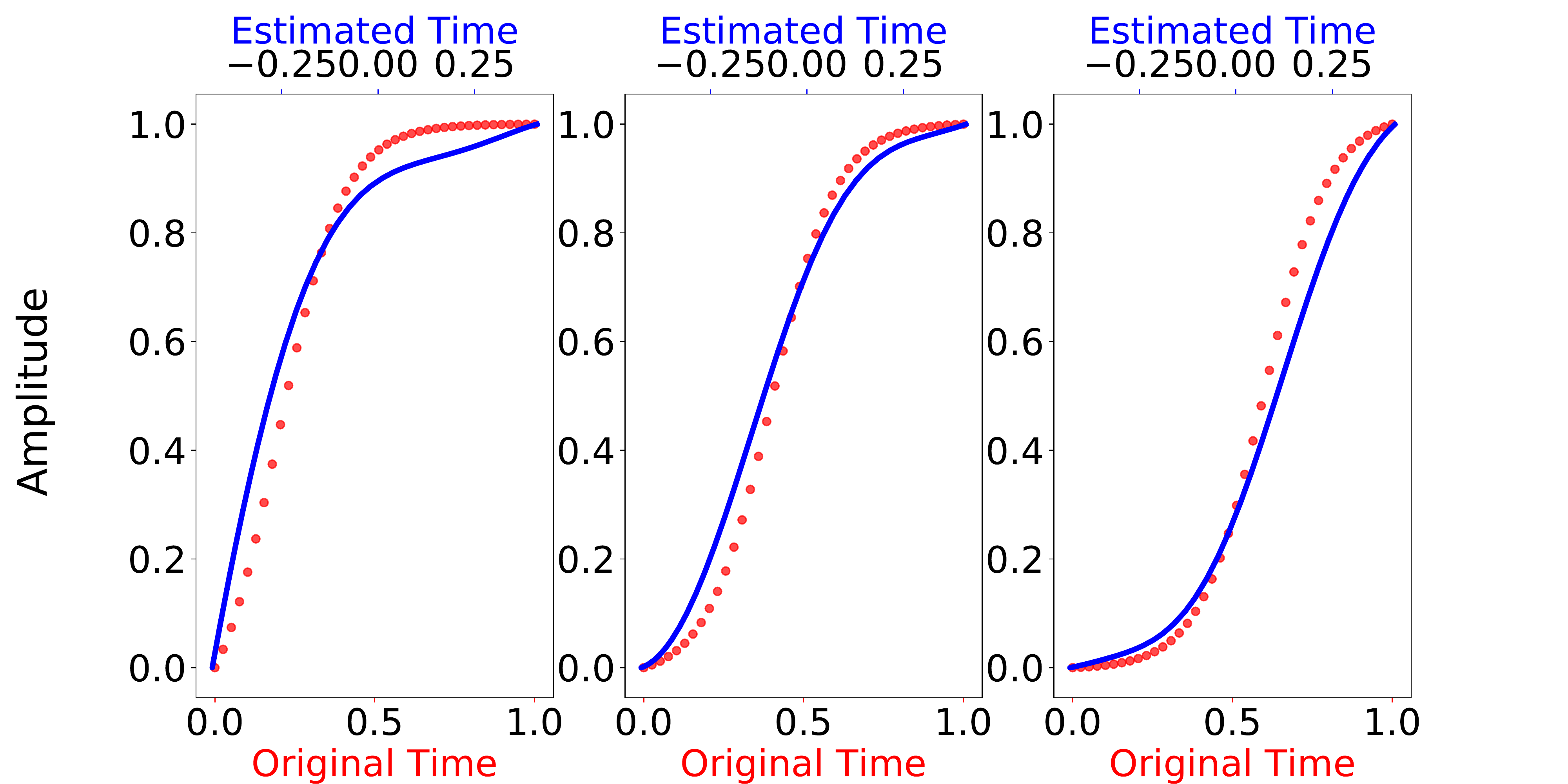}
    \caption{}
    \label{fig:estimated_S}
  \end{subfigure}
  \begin{subfigure}[c]{0.55\textwidth}
    \includegraphics[width=\textwidth]{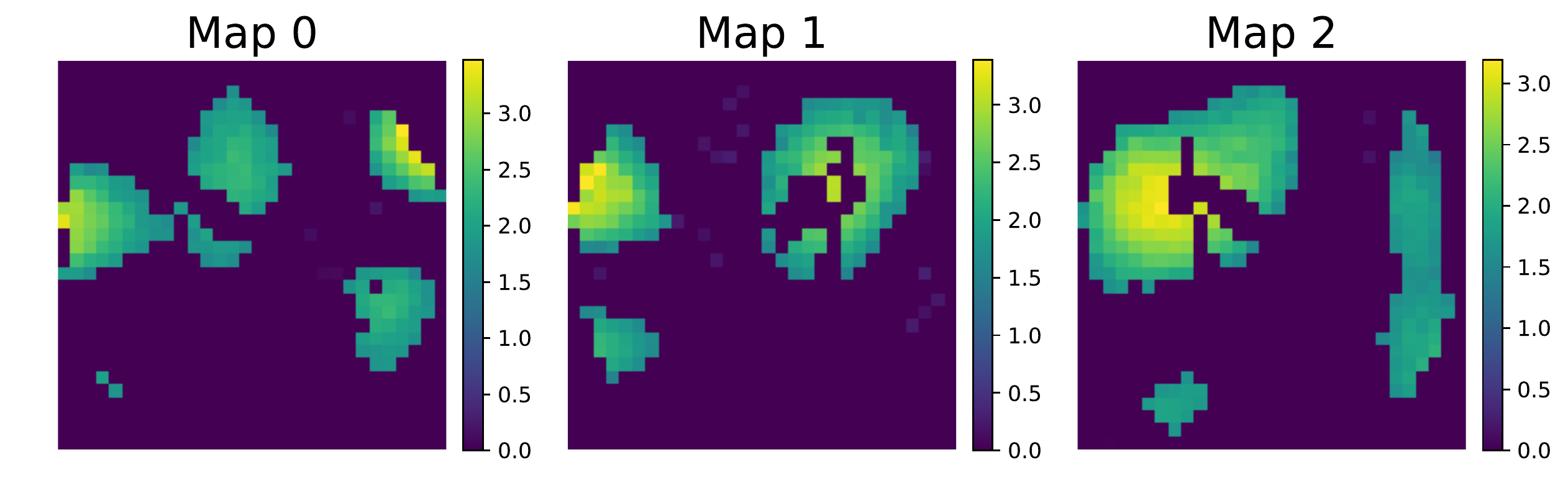}
    \caption{}
    \label{fig:estimated_A}
\end{subfigure}
\caption{(a)-(b) Ground truth temporal and spatial sources. (c) Red : raw temporal sources against the original time axis. Blue : recovered temporal sources against the estimated time scale. (d) Estimated spatial maps.}
\label{fig:original_sources}
\end{figure}

\begin{figure}
  \centering
  \includegraphics[width=0.30\textwidth]{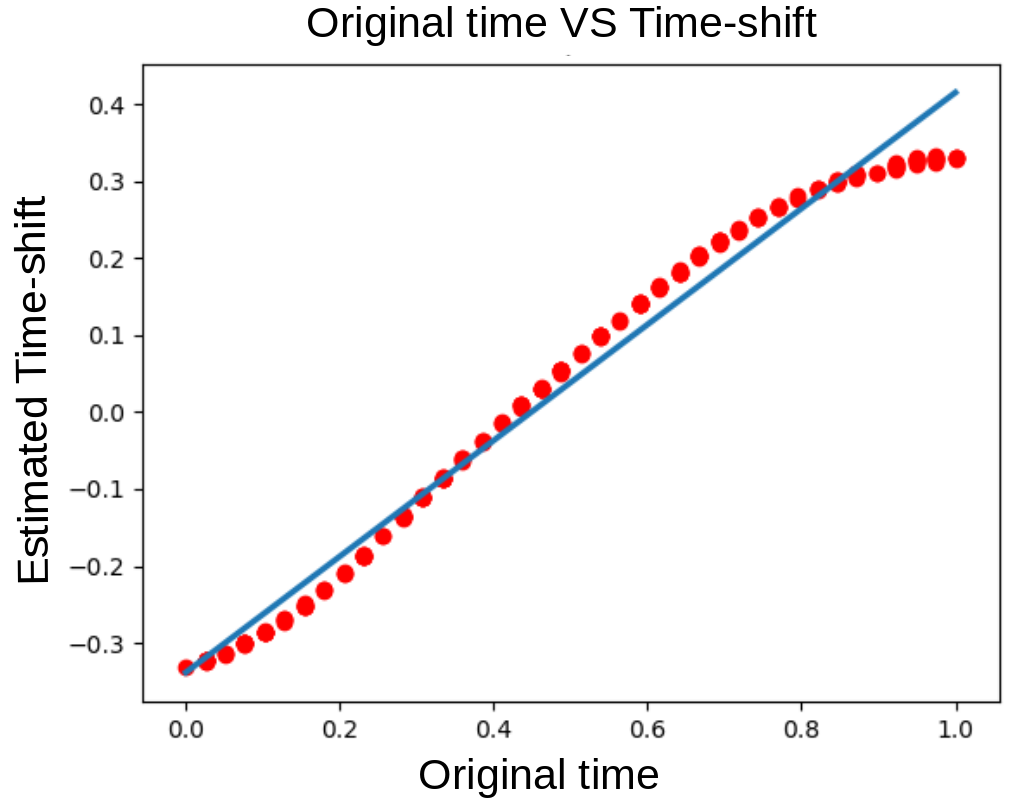}
  \centering
  \caption{The red points represent the values of the estimated subjects' time-shift against their associated ground truth value.}
  \label{fig:estimated_time}
\end{figure}

\begin{figure}
  \begin{subfigure}[c]{0.26\textwidth}
    \includegraphics[width=\textwidth]{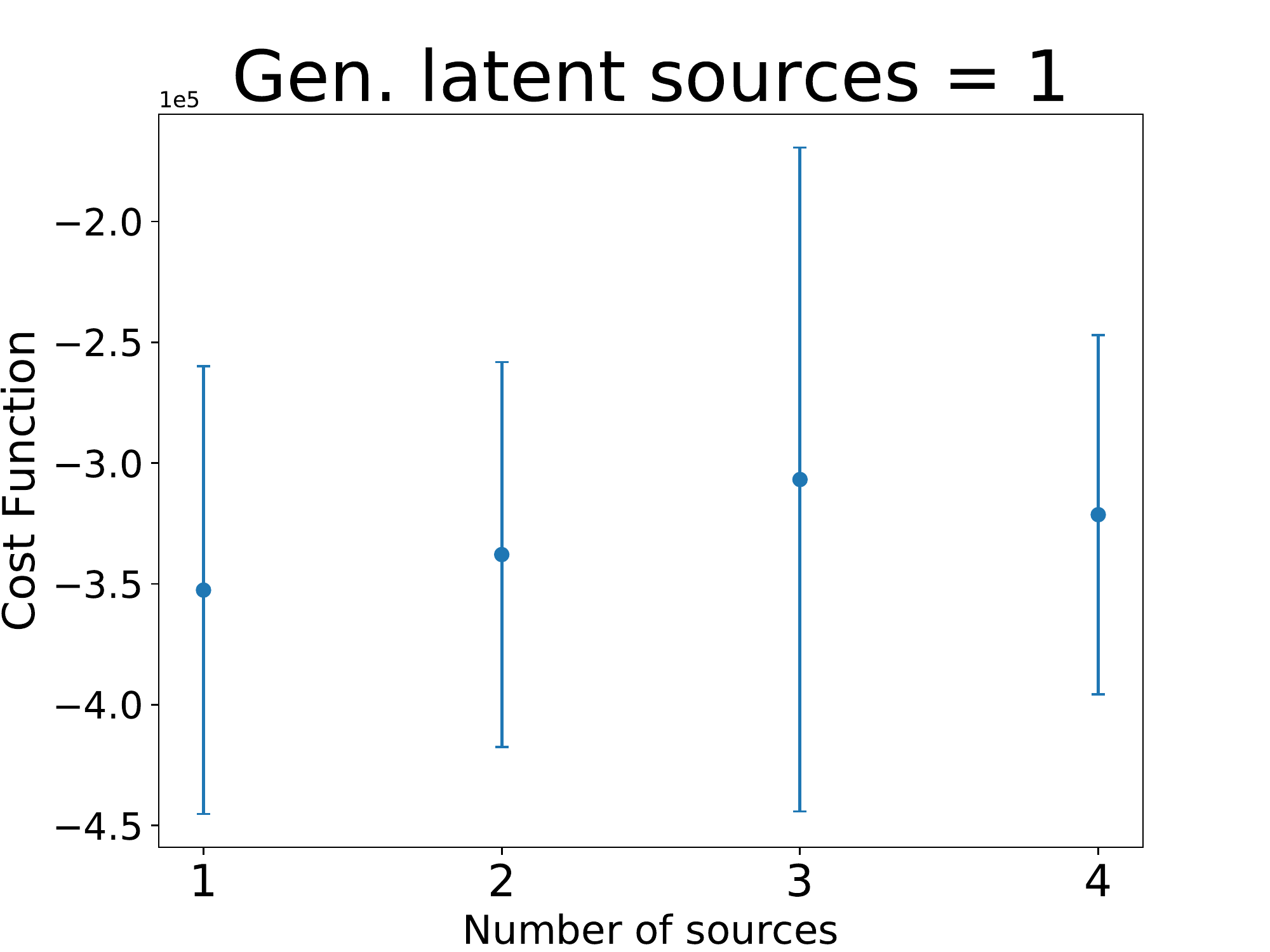}
    \caption{}
    \label{fig:1source}
  \end{subfigure}
  \begin{subfigure}[c]{0.26\textwidth}
    \includegraphics[width=\textwidth]{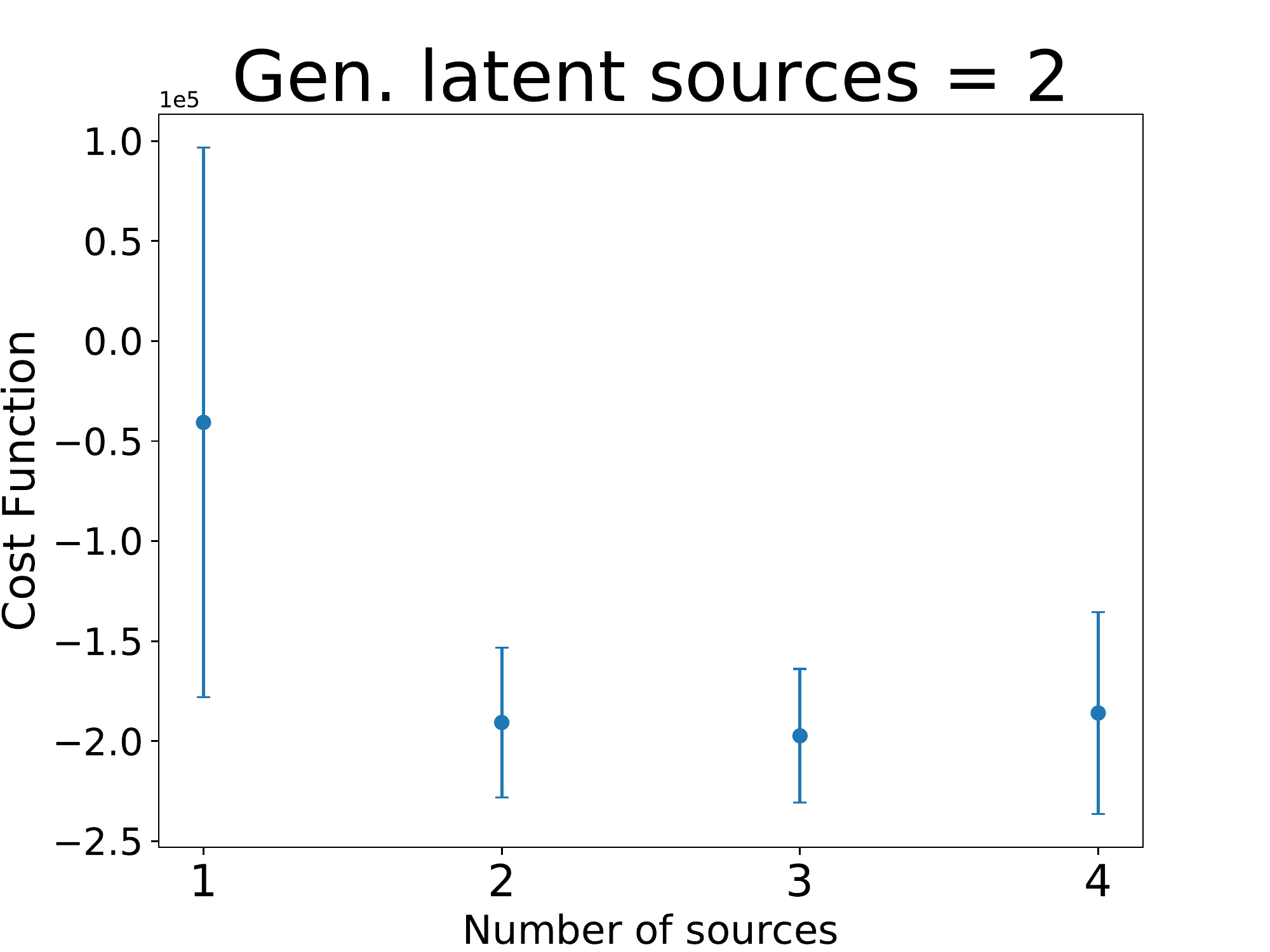}
    \caption{}
    \label{fig:2sources}
  \end{subfigure}
    \begin{subfigure}[c]{0.26\textwidth}
    \includegraphics[width=\textwidth]{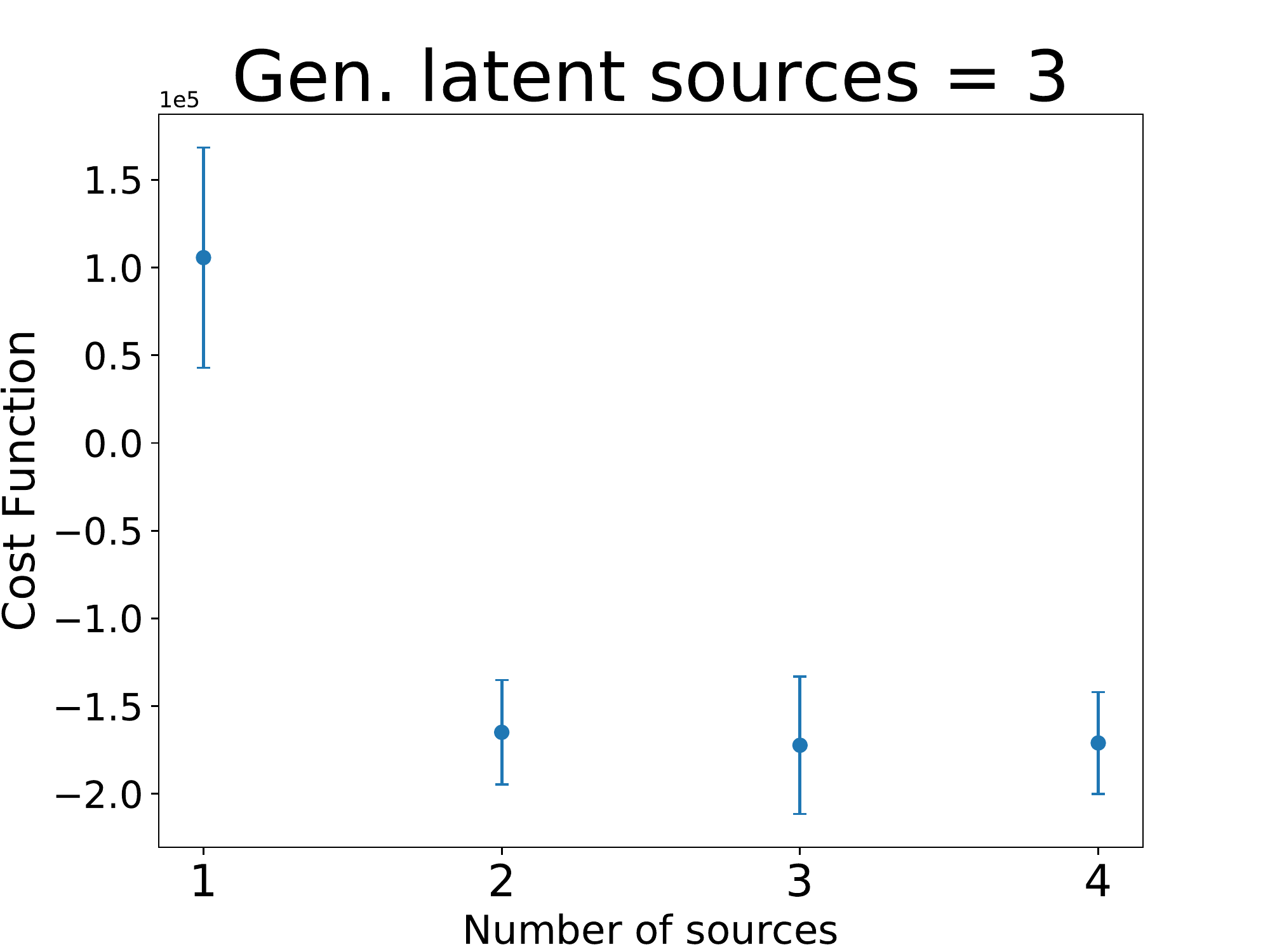}
    \caption{}
    \label{fig:3sources}
  \end{subfigure}
    \begin{subfigure}[c]{0.19\textwidth}
    \includegraphics[width=\textwidth]{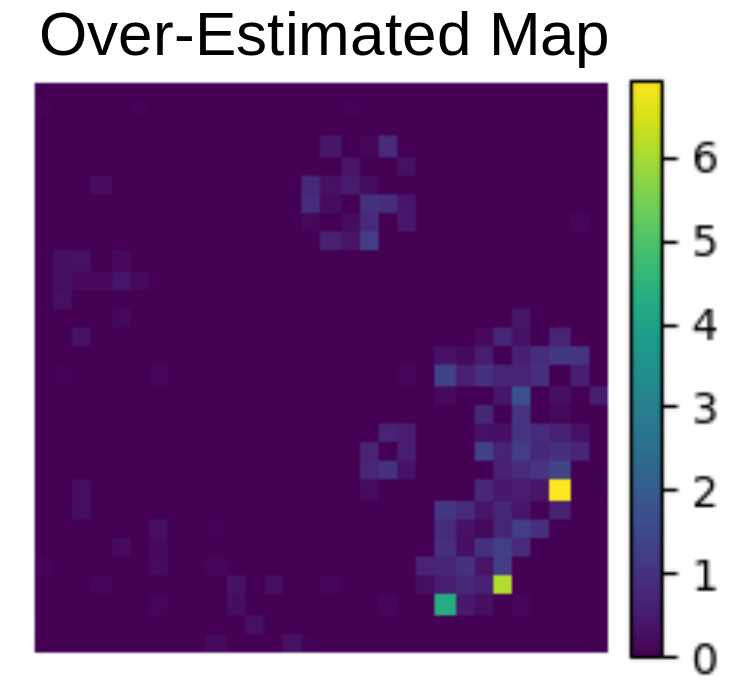}
    \caption{}
    \label{fig:zero_map}
  \end{subfigure}
\caption{(a)-(b)-(c) : Distribution of the lower bound against the number of fitted sources. (d) : $4^{th}$ extracted spatial map with data generated by 3 latent sources.}  
\label{fig:model_selection}
\end{figure}

We tested the algorithm on synthetic data to assess its ability to separate spatio-temporal sources from mixed data, and to provide a model selection via the variational lower bound. We generated three monotonically increasing functions $\vect{S}_{i}(t)$ such that $\vect{S}_i(t) = 1/(1 + \exp(-t + \alpha_{i}))$, and three synthetic Gausian activation maps $\vect{A}_1, \vect{A}_2, \vect{A}_3$ with a $30 \times 30$ resolution, to mimick grey matter brain areas (Figures \ref{fig:synthetic_S} and \ref{fig:synthetic_A}). 
\noindent
The data was generated as $\vect{Y}_{p,j} = \vect{S}(t_{p})\vect{A} + \vect{\mathcal{E}}_{j}$ over $40$ time points $t_{p}$, where $t_{p}$ is uniformly distributed in [0,1]. We sampled $50$ images at instants $t_{p}$ and applied our method. To simulate a pure cross-sectional setting the time associated to each input image was set to zero. Figures \ref{fig:estimated_S} and \ref{fig:estimated_A} show the estimated spatio-temporal processes when fitting the model with three latent sources.
In Figure \ref{fig:estimated_time}, we see that the individual time-shift parameter estimated for each subject correlates with the original time used to generate the data. This means that the algorithm correctly positions each subject on the temporal trajectories.
\newline
\newline
To test the model selection, we generated the data as described above using respectively one, two, or three sources over ten folds. For each fold we ran the algorithm looking for one to four sources. Figure \ref{fig:model_selection} shows mean and standard deviation of the lower bound. We observe that when the number of sources is under-estimated the lower bound is higher. When the number of sources is over-estimated, although the lower bound for model selection is more uncertain, by looking at the extracted spatial maps we observe that the additional sources are mainly set to zero or have low weights (see the map of Figure \ref{fig:model_selection}). These experimental results indicate that the optimal number of sources should be selected by inspection of both the lower bound and the extracted spatial sources.
\clearpage
\subsection{Comparison with ICA}
\begin{figure}
  \begin{subfigure}[c]{\textwidth}
    \includegraphics[width=0.44\textwidth]{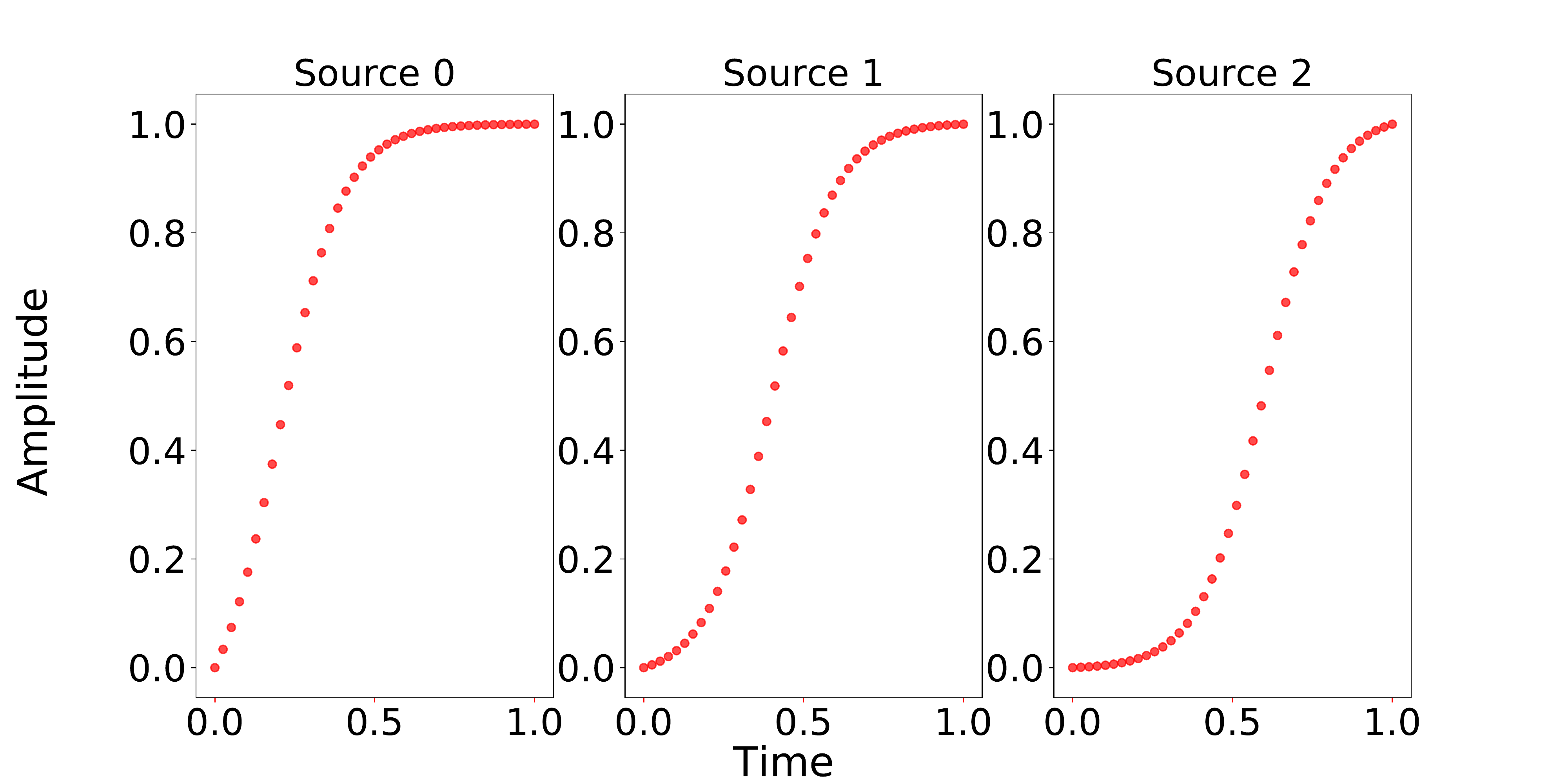}
    \includegraphics[width=0.55\textwidth]{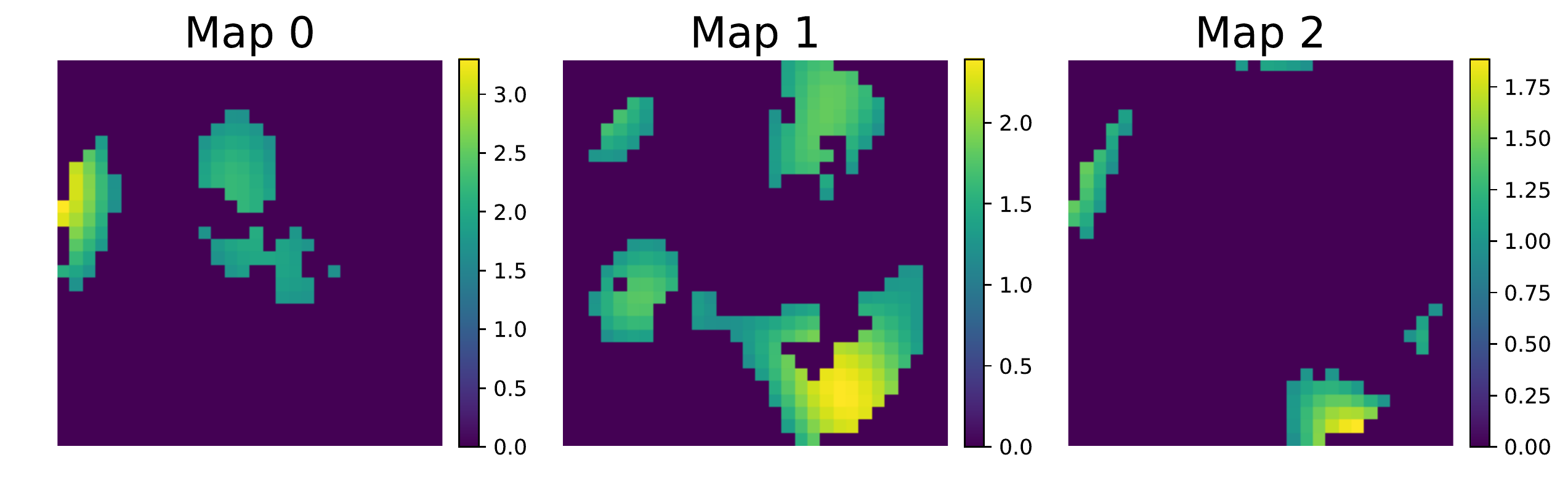}
        \caption{Ground truth}
    \label{fig:original}
  \end{subfigure}
    \begin{subfigure}[c]{\textwidth}
    \includegraphics[width=0.44\textwidth]{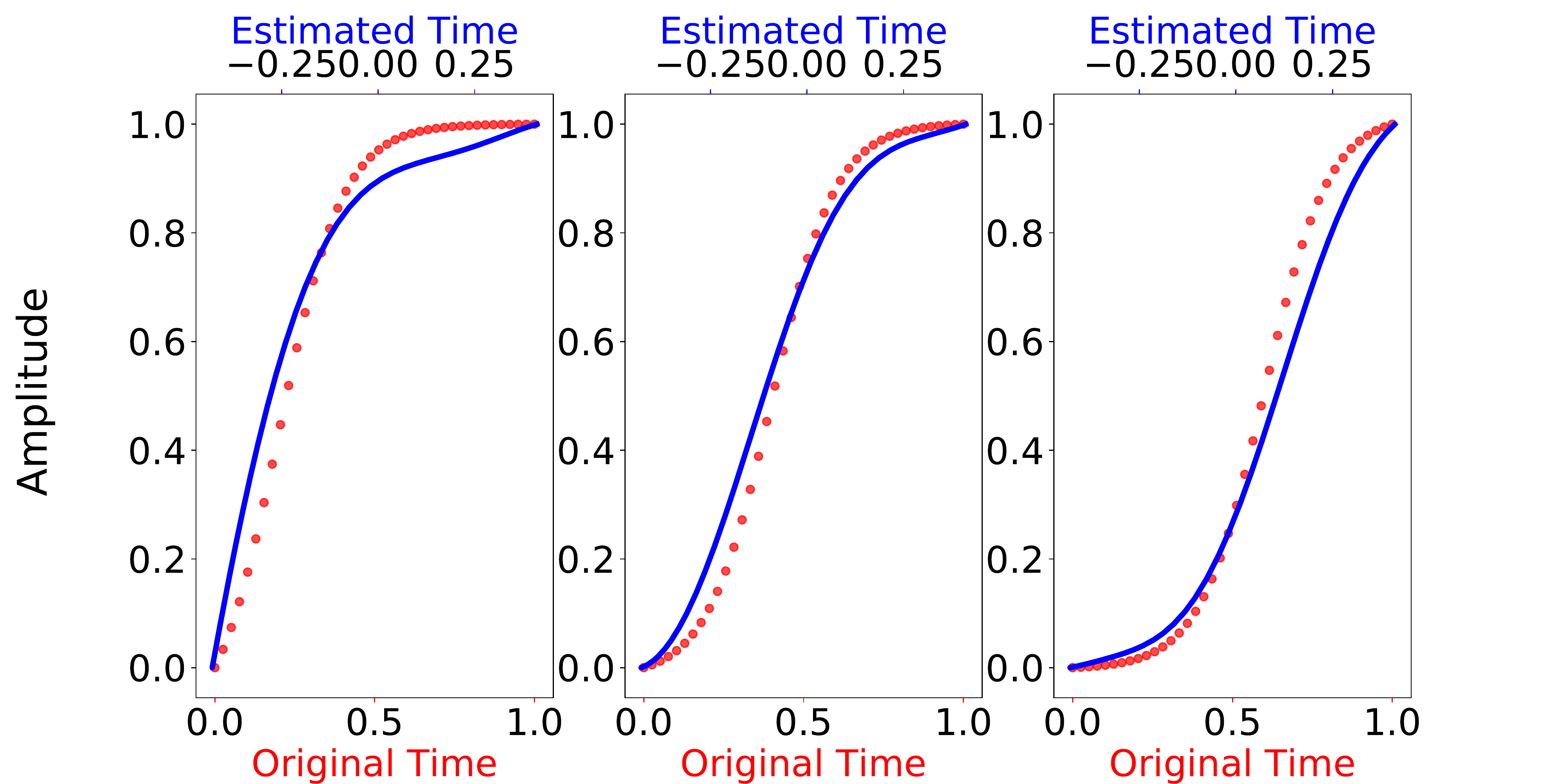}
    \includegraphics[width=0.55\textwidth]{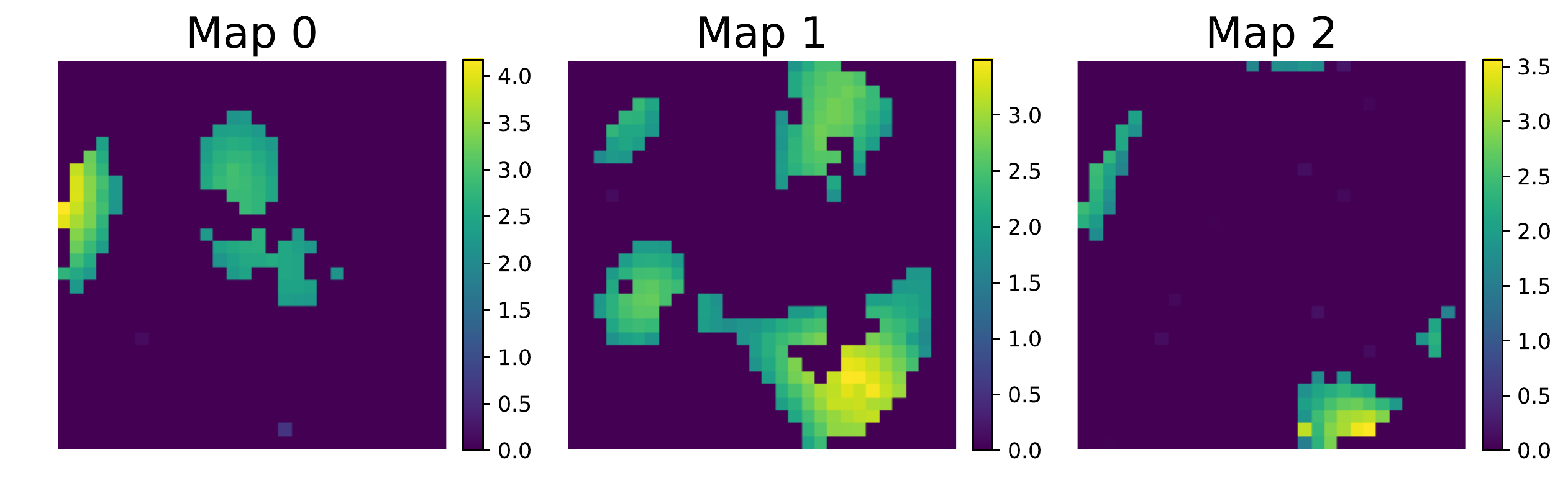}
        \caption{IGPA estimation}
    \label{fig:igpa}
  \end{subfigure}
    \begin{subfigure}[c]{\textwidth}
    \includegraphics[width=0.44\textwidth]{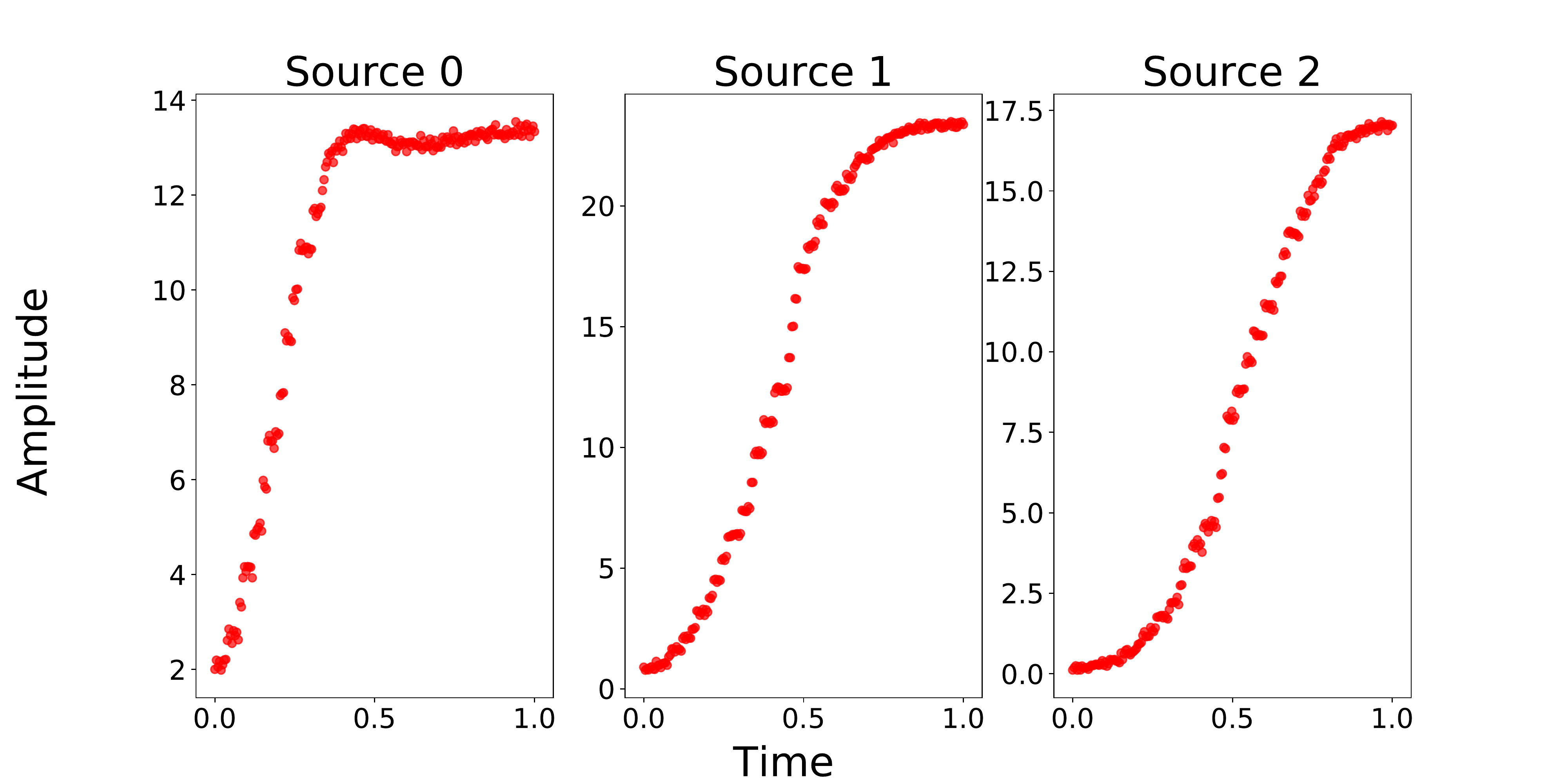}
    \includegraphics[width=0.55\textwidth]{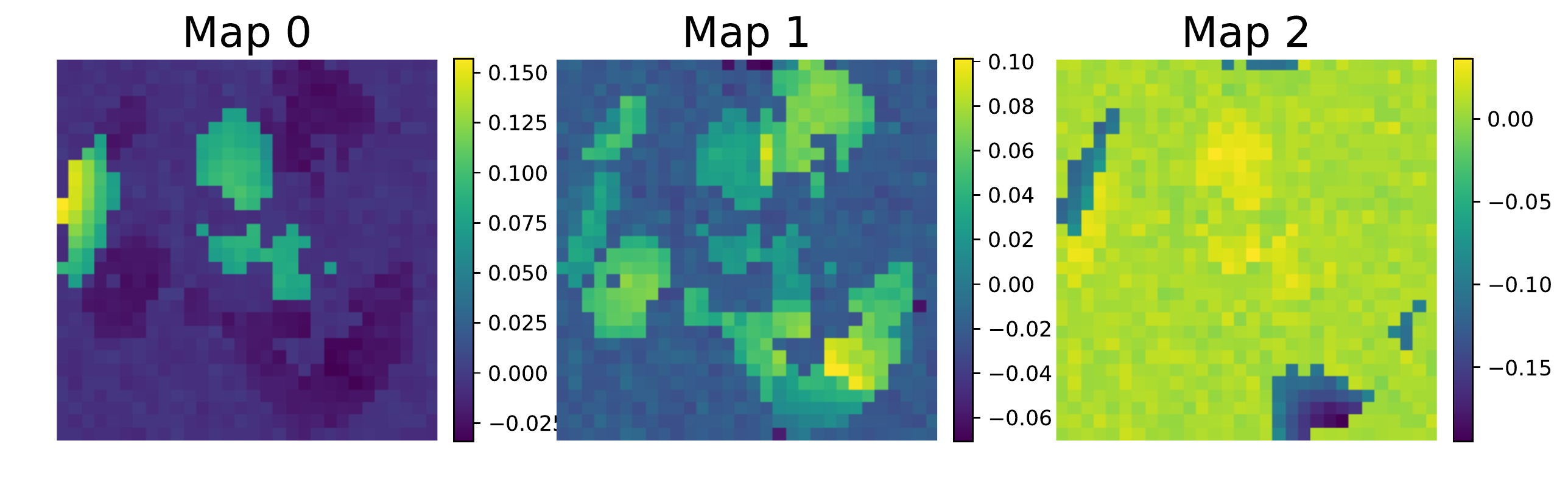}
        \caption{ICA estimation}
    \label{fig:ica}
    \end{subfigure}   %
\caption{Ground truth and estimations of the spatio-temporal processes given by both our method and ICA.}
  \label{fig:ica_comparison}
\end{figure}

We performed a comparison of our algorithm with ICA on an example similar to the one of section 3.1. However, since standard ICA can't be applied when the time associated to each image is unknown, the data was generated in a simplified setting. To do so we assigned the ground truth parameter $t_{p}$ beforehand. The goal was to compare the separation performances of both our algorithm and ICA, on data generated with three latent spatio-temporal processes. In Figure \ref{fig:ica_comparison} we observe that the sources estimated by ICA are more noisy and uncertain than the ones estimated by our method, highlighting the importance of the priors and constraints introduced in our model.
\vspace{-5mm}
\subsection{Application on Real Data}
\label{sec:real_data}
\begin{figure}
  \begin{subfigure}[c]{0.50\textwidth}
    \includegraphics[width=\textwidth]{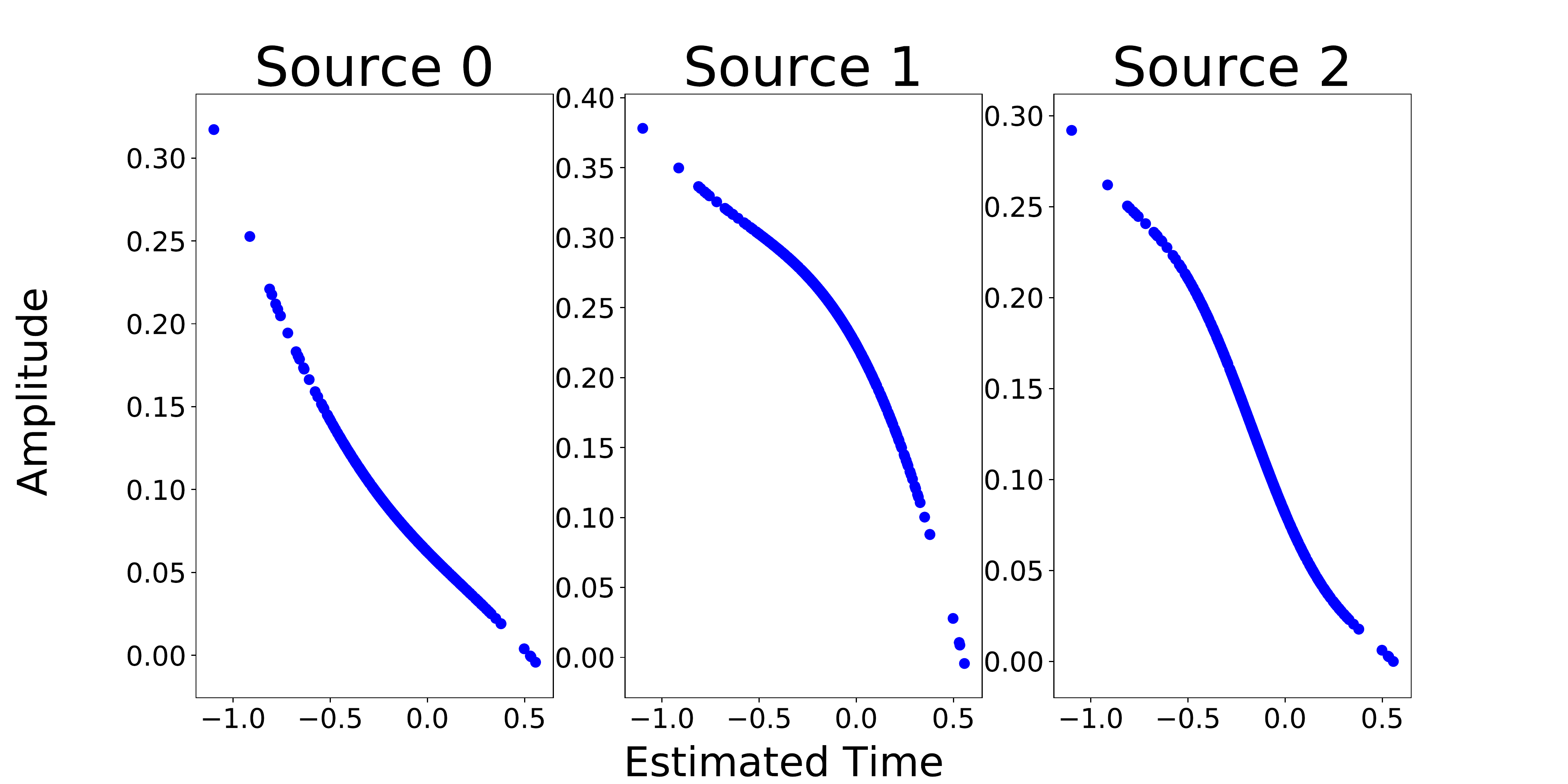}
    \caption{}
    \label{fig:real_data_S}
  \end{subfigure}
  \begin{subfigure}[c]{0.49\textwidth}
    \includegraphics[width=\textwidth]{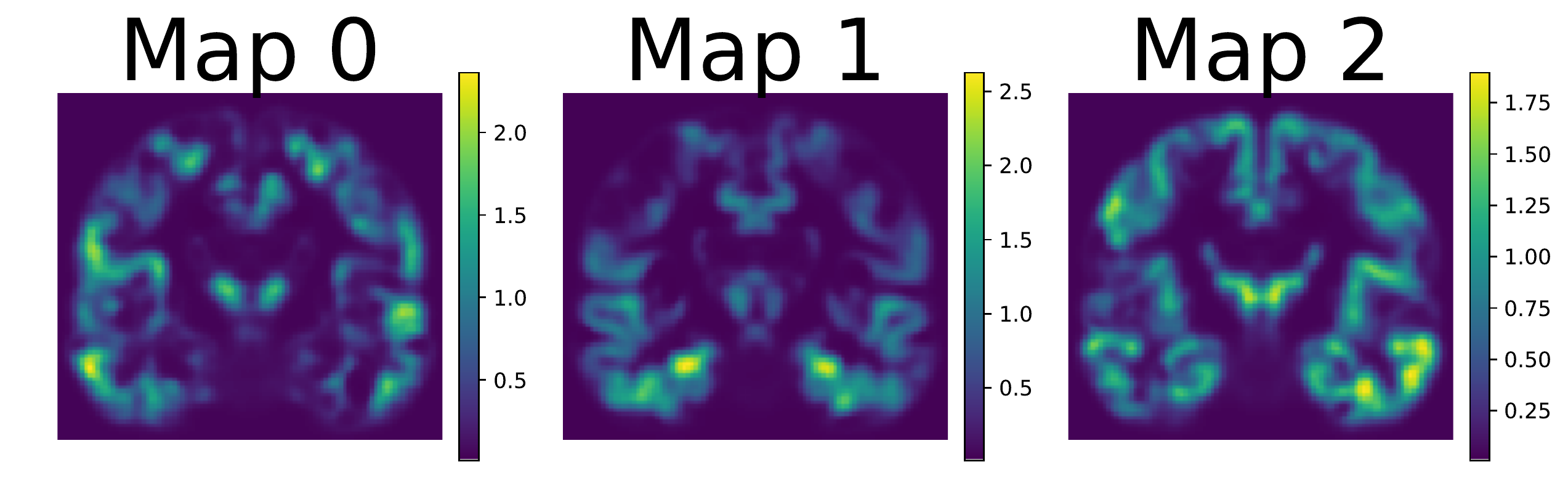}
    \caption{}
    \label{fig:real_data_A}
  \end{subfigure}
  \caption{(a)-(b) Temporal and Spatial sources extracted from the data.}
\label{fig:estimated_real_sources}
\end{figure}

\begin{figure}
  \begin{subfigure}[c]{0.24\textwidth}
    \includegraphics[width=\textwidth]{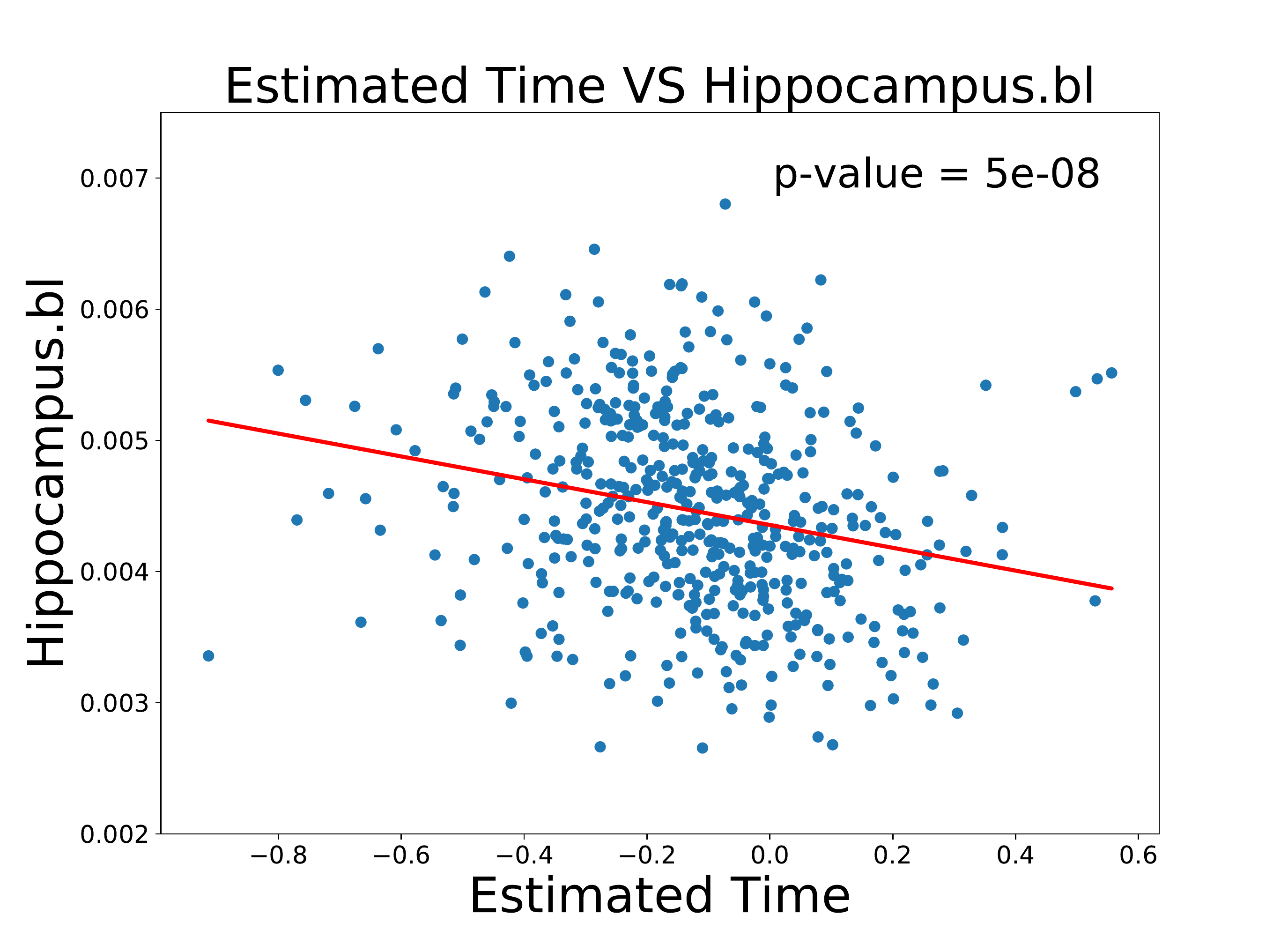}
    \caption{}
    \label{fig:hippocampus}
  \end{subfigure}
  \begin{subfigure}[c]{0.24\textwidth}
    \includegraphics[width=\textwidth]{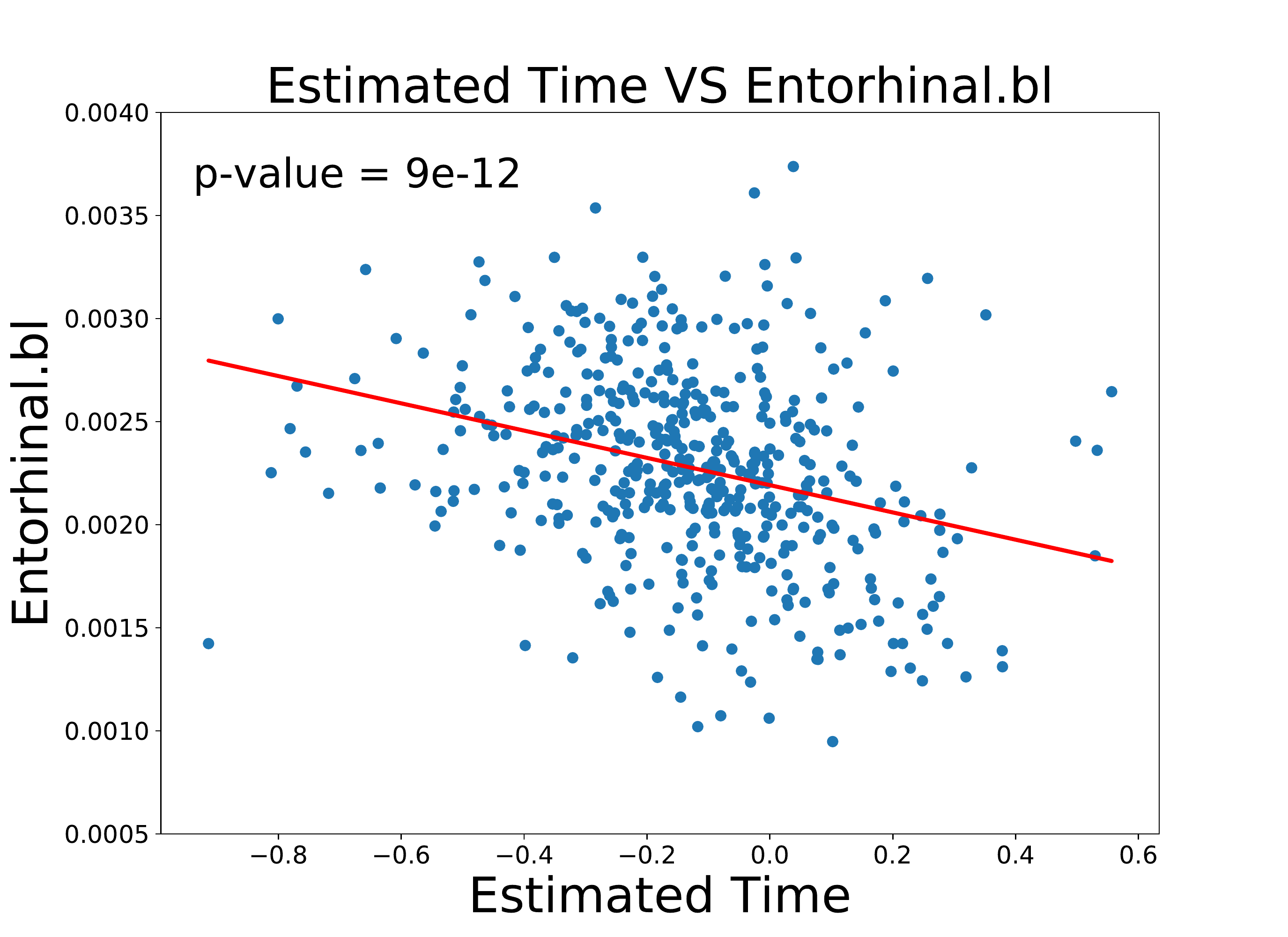}
    \caption{}
    \label{fig:entorhinal}
  \end{subfigure}
    \begin{subfigure}[c]{0.24\textwidth}
    \includegraphics[width=\textwidth]{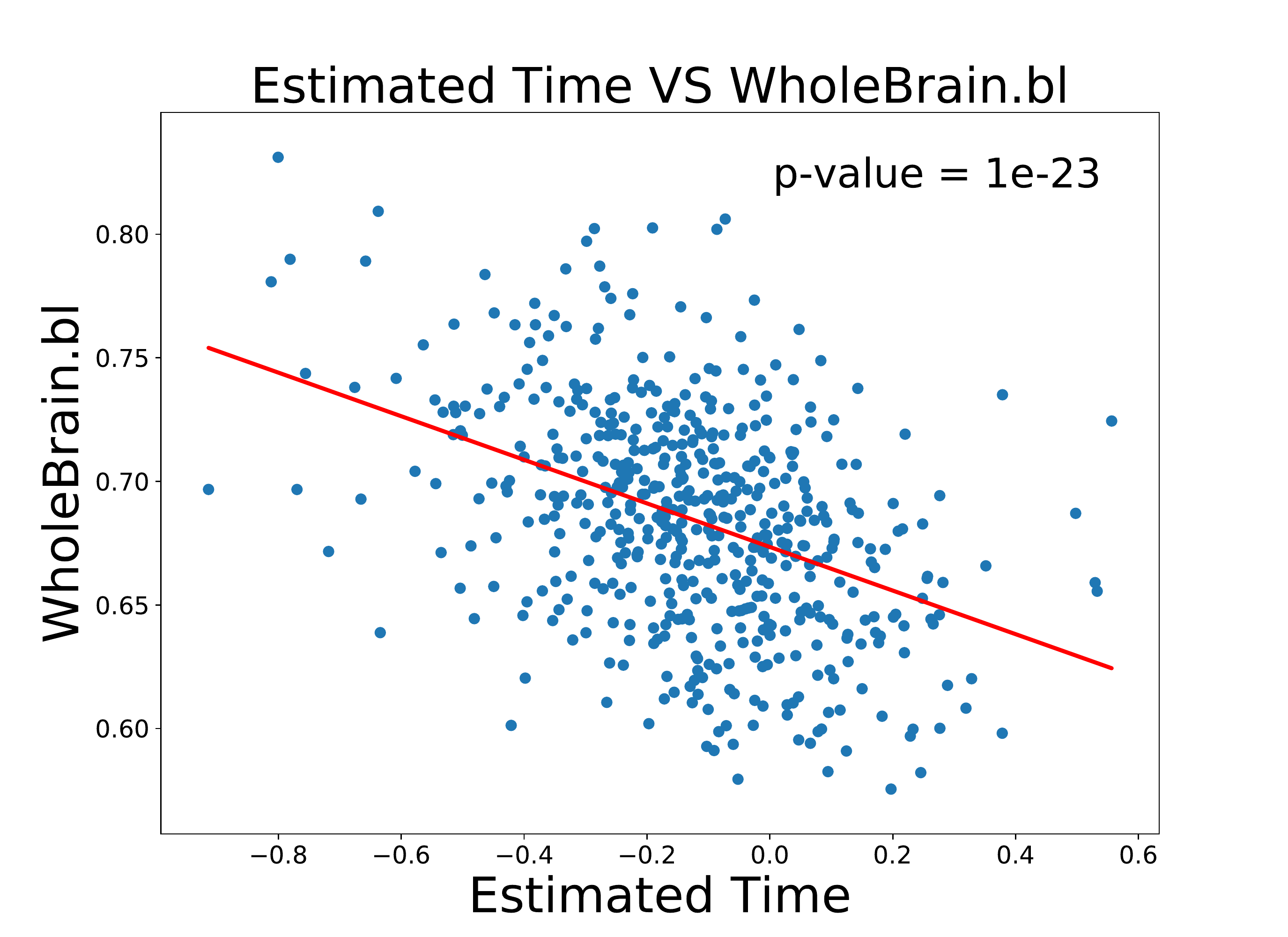}
    \caption{}
    \label{fig:wholebrain}
  \end{subfigure}
    \begin{subfigure}[c]{0.24\textwidth}
    \includegraphics[width=\textwidth]{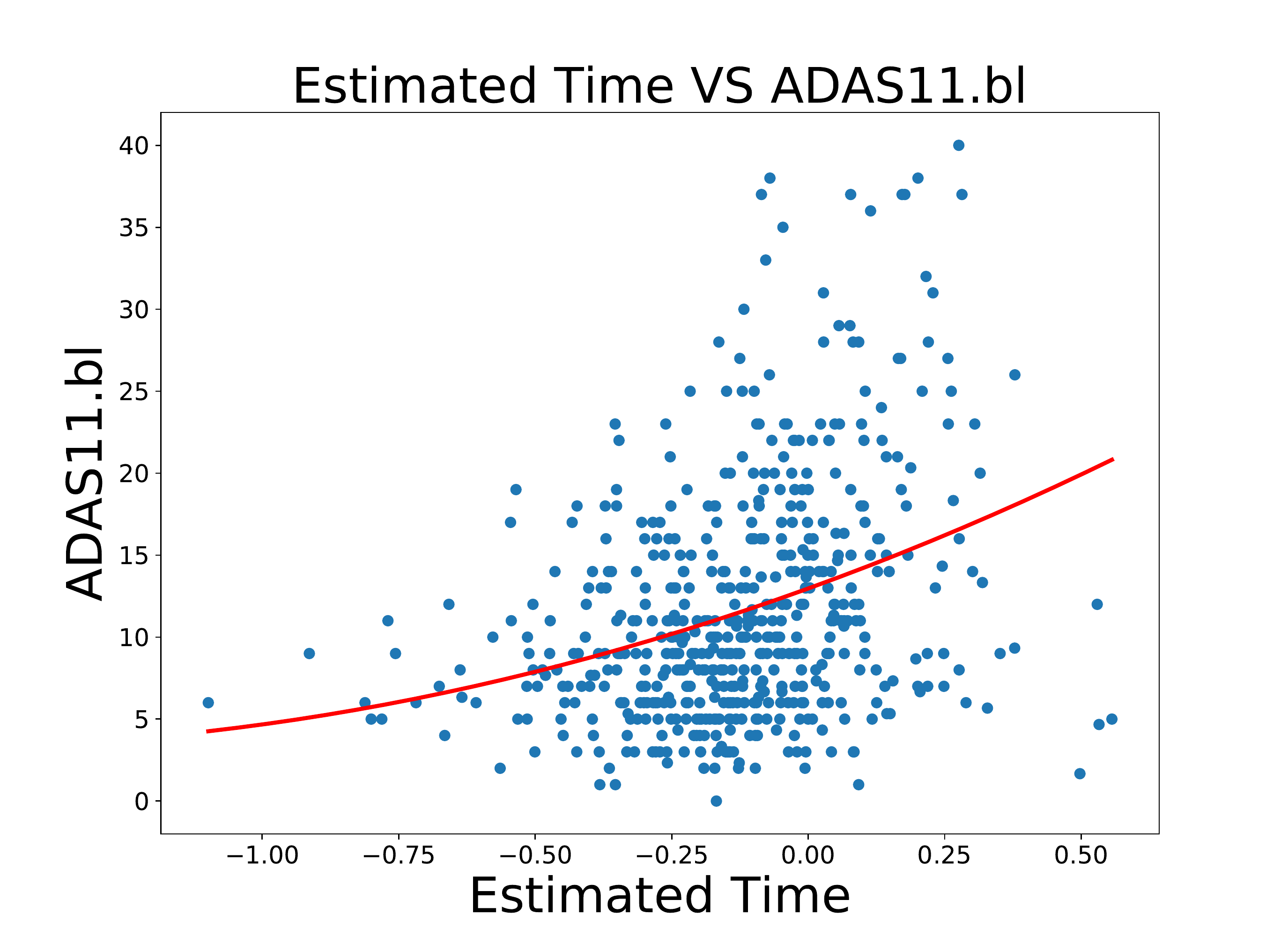}
    \caption{}
    \label{fig:adas11}
  \end{subfigure}
  \caption{Evolution of volumetric and clinical biomarkers along the estimated time.}
\label{fig:real_data_corr}
\end{figure}

Data  used in the preparation of this article were obtained from the Alzheimer’s Disease Neuroimaging Initiative (ADNI) database (adni.loni.usc.edu). The ADNI was launched in 2003 as  a public-private partnership, led by  Principal Investigator Michael W. Weiner, MD. For up-to-date information, see www.adni-info.org.
\newline
\newline
\noindent
In this section we present an application of the algorithm on real data, using grey matter maps extracted from structural MRI. We selected a cohort of 555 subjects from ADNI composed by 94 healthy controls, 343 MCI, and 118 AD patients. We processed the baseline MRI of each subject to obtain high-dimensional grey matter density maps in a standard space \cite{ref_dartel}. We extracted the $90 \times 100$ middle coronal slice for each patient, to obtain a data matrix $Y$ with dimensions $555 \times 9000$, and applied our algorithm looking for three spatio-temporal sources (see Figure \ref{fig:estimated_real_sources}).
\noindent
The middle spatial map shows a strong activation of the hippocampus, while the left and right plots show an activation on the temporal lobes, with two similar temporal behaviours, characterized by a less pronounced grey matter loss compared to the hippocampus. More specifically, we observe that the hippocampal trajectory has a strong acceleration in opposition to the other brain areas. This pattern quantified by our model in a pure data-driven manner is compatible with empirical evidence from clinical studies \cite{ref_biomarker_changes}. In Figure \ref{fig:real_data_corr} we observe the estimated time of each patient against standard volumetric and clinical biomarkers. We see a strong correlation between brain volumetric measures and the estimated time, as well as a non-linear relation in the evolution of ADAS11. The latter result indicates an acceleration of clinical symptoms along the estimated time course.

\section{Conclusion}

We presented a method for analyzing spatio-temporal data, which provides both independent spatio-temporal processes at stake in AD, and a disease progression scale. Applied on grey matter maps, the model highlights different brain regions affected by the disease, such as the hippocampus and the temporal lobes, along with their differential temporal trajectory. We also show a strong correlation between the estimated disease progression scale and different clinical and volumetric biomarkers. We are currently extending the approach to scale to 3D volumetric images by parallelization on multiple GPUs. The lower bound properties will be also further investigated to better assess its reliability, in order to improve the model comparison. Moreover the method will be extended beyond the cross-sectional application of section \ref{sec:real_data}, to account for time-series of brain images, as well as for multimodal imaging biomarkers. Finally we will investigate the use of the approach for prognosis purposes, to provide a data-driven assessment of disease severity in testing patients.

\section{Acknowledgements}
This work has been supported by the French government, through the UCA\textsuperscript{JEDI} Investments in the Future project managed by the National Research Agency (ref.n ANR-15-IDEX-01), the grant AAP Santé 06 2017-260 DGA-DSH, and by the Inria Sophia Antipolis - Méditerranée, "NEF" computation cluster.

\bibliographystyle{splncs04}

\clearpage
\section*{\centerline{Appendix}}
\setcounter{equation}{0}

\subsection*{A. Lower bound derivation}
In this section we detail the derivation of the lower bound :
\begin{align} 
\begin{split}
\label{eq:final_cost}
\log(p(\vect{Y}, \mathcal{C}|\sigma, \lambda)) & = \log[\int_{\vect{A}}\int_{\vect{S}}p(\vect{Y}|\vect{A},\vect{S}, \sigma)p(\mathcal{C}|\vect{S'}, \lambda)p(\vect{A})p(\vect{S})d\vect{A} d\vect{S}] \\
& =\log[\int_{\vect{A}}\int_{\vect{\Omega}}\int_{\vect{W}}p(\vect{Y}|\vect{A}, \vect{\Omega}, \vect{W}, \sigma)p(\mathcal{C}|\vect{\Omega}, \vect{W}, \lambda)p(\vect{A})p(\vect{\Omega}, \vect{W})d\vect{A} d\vect{\Omega} \vect{dW}] \\
& =\log[\int_{\vect{A}}\int_{\vect{\Omega}}\int_{\vect{W}}p(\vect{Y}|\vect{A},\vect{\Omega}, \vect{W}, \sigma)p(\mathcal{C}|\vect{\Omega}, \vect{W}, \lambda)p(\vect{A})p(\vect{\Omega})p(\vect{W})d\vect{A} d\vect{\Omega} \vect{dW}] \\
& = \log[\int_{\vect{A}}\int_{\vect{\Omega}}\int_{\vect{W}}p(\vect{Y}|\vect{A},\vect{\Omega}, \vect{W}, \sigma)p(\mathcal{C}|\vect{\Omega}, \vect{W}, \lambda)p(\vect{A})p(\vect{\Omega})p(\vect{W})\frac{q_{1}(\vect{A})q_{2}(\vect{\Omega})q_{3}(\vect{W})}{q_{1}(\vect{A})q_{2}(\vect{\Omega})q_{3}(\vect{W})}d\vect{A} d\vect{\Omega} \vect{dW}] \\
& = \log[\E_{\vect{A} \sim q_{1}, \vect{\Omega} \sim q_{2}, \vect{W} \sim q_{3}}[\frac{p(\vect{Y}|\vect{A},\vect{\Omega}, \vect{W}, \sigma)p(\mathcal{C}|\vect{\Omega}, \vect{W}, \lambda)p(\vect{A})p(\vect{\Omega})p(\vect{W})}{q_{1}(\vect{A})q_{2}(\vect{\Omega})q_{3}(\vect{W})}]] \\
& \geqslant \E_{\vect{A} \sim q_{1}, \vect{\Omega} \sim q_{2}, \vect{W} \sim q_{3}}[\log[\frac{p(\vect{Y}|\vect{A},\vect{\Omega}, \vect{W}, \sigma)p(\mathcal{C}|\vect{\Omega}, \vect{W}, \lambda)p(\vect{A})p(\vect{\Omega})p(\vect{W})}{q_{1}(\vect{A})q_{2}(\vect{\Omega})q_{3}(\vect{W})}]] \\
& =  \E_{\vect{A} \sim q_{1}, \vect{\Omega} \sim q_{2}, \vect{W} \sim q_{3}}[\log[p(\vect{Y}|\vect{A},\vect{\Omega}, \vect{W}, \sigma)]] + \E_{\vect{\Omega} \sim q_{2}, \vect{W} \sim q_{3}}[log[p(\mathcal{C}|\vect{\Omega}, \vect{W}, \lambda)]] \\
& - \mathcal{D}[q_{1}(\vect{A})||p(\vect{A})] - \mathcal{D}[q_{2}(\vect{\Omega})||p(\vect{\Omega})] - \mathcal{D}[q_{3}(\vect{W})||p(\vect{W})].
\end{split}
\end{align}
In the Method section we introduced the approximation $q_{1}(\vect{A}) =  \prod_{n=1}^{Ns} \mathcal{N}(\vect{\mu_{n}}, \vect{\Sigma}(\alpha, \beta))$. The covariance matrix is shared by all the spatial processes which gives us the set of spatial parameters : \begin{equation} \psi = \{\vect{\mu_{n}}, n \in [1, Ns], \alpha, \beta\}. \end{equation} 
We defined the approximated distributions $q_{2}(\vect{\Omega}) = \prod_{n,j} \mathcal{N}(r_{n,j}, p_{n,j}^{2})$ and $q_{3}(\vect{W}) = \prod_{n,j} \mathcal{N}(m_{n,j}, s_{n,j}^{2})$, leading to the set of temporal parameters :
\begin{equation} \theta = \{\vect{m_{n}}, \vect{s_{n}}, \vect{r_{n}}, \vect{p_{n}}, l_{n}, n \in [1, Ns]\}. \end{equation}
Now we can obtain every term of \eqref{eq:final_cost}.
The Kullback-Leibler of a multivariate Gaussian has a closed-from : 
\begin{align}
\mathcal{D}[q_{1}(\vect{A})||p(\vect{A})] & = 
\frac{1}{2}\sum_{n=1}^{Ns}Tr(\vect{\Sigma}) + \vect{\mu_{n}}^{T}\vect{\mu_{n}} - F - \log[det(\vect{\Sigma})]. \\
\mathcal{D}[q_{2}(\vect{\Omega})|p(\vect{\Omega})] & = 
\frac{1}{2} \displaystyle \sum_{n=1}^{Ns} \sum_{j} \vect{p}_{n,j}^{2}l_{n} + \vect{r}_{n,j}^{2}l_{n}, - 1 - log(\vect{p}_{n,j}^{2}l_{n}). \\
\mathcal{D}[q_{3}(\vect{W})|p(\vect{W})] & = \frac{1}{2} \displaystyle \sum_{n=1}^{Ns} \sum_{j} \vect{s}_{n,j}^{2} + \vect{m}_{n,j}^{2} - 1 - log(\vect{s}_{n,j}^{2}).
\end{align}
As in \cite{ref_kingma} we employ the reparameterization trick to have an efficient way of sampling the expectations of \eqref{eq:final_cost}. Thus we have :

        \begin{itemize}
        \item $\vect{\Omega}_{n,j} = \vect{r}_{n,j} + \vect{p}_{n,j}*\vect{\zeta}_{n,j}$,
        \item $\vect{W}_{n,j} = \vect{m}_{n,j} + \vect{s}_{n,j}*\vect{\epsilon}_{n,j}$,
        \item $\vect{A_{n}} = \vect{\mu_{n}} + \vect{\Sigma}^{\frac{1}{2}}*\vect{\kappa}$,
        \end{itemize}
Which gives us :
\begin{align}
\E_{\vect{A} \sim q_{1}, \vect{\Omega} \sim q_{2}, \vect{W} \sim q_{3}}[log(p(\vect{Y}|\vect{A},\vect{\Omega}, \vect{W}, \sigma))] & =  \E_{\vect{\epsilon}, \vect{\zeta}, \vect{\kappa}}[\log(p(\vect{Y}|\vect{m},\vect{s}, \vect{r}, \vect{p}, \vect{\mu}, \vect{\Sigma}, \sigma))], \\
\E_{\vect{\Omega} \sim q_{2}, \vect{W} \sim q_{3}}[log(p(\mathcal{C}|\vect{\Omega}, \vect{W}, \lambda))] & =  \E_{\vect{\epsilon}, \vect{\zeta}}[log(p(\mathcal{C}|\vect{m},\vect{s}, \vect{r}, \vect{p}, \lambda))].
\end{align}
Where $\vect{\zeta_{n}}$, $\vect{\epsilon_{n}}$ and $\vect{\kappa}$ follow a $ \mathcal{N}(0,\vect{I})$ distribution.
\clearpage
\subsection*{B. ADNI}
Data collection and sharing for this project was funded by the Alzheimer's Disease Neuroimaging Initiative (ADNI) and DOD ADNI. ADNI is funded by the National Institute on Aging, the National Institute of Biomedical Imaging and Bioengineering, and through generous contributions from the following: AbbVie, Alzheimer’s Association; Alzheimer’s Drug Discovery Foundation; Araclon Biotech; BioClinica, Inc.; Biogen; Bristol-Myers Squibb Company;CereSpir,  Inc.;Cogstate;Eisai Inc.; Elan Pharmaceuticals, Inc.; Eli Lilly and Company; EuroImmun; F. Hoffmann-La Roche Ltd and its  affiliated  company  Genentech, Inc.;  Fujirebio;  GE  Healthcare; IXICO  Ltd.; Janssen Alzheimer Immunotherapy Research \& Development, LLC.; Johnson \& Johnson Pharmaceutical Research \& Development LLC.;Lumosity;Lundbeck;Merck \& Co., Inc.; Meso Scale Diagnostics, LLC.;NeuroRx Research; Neurotrack Technologies;Novartis Pharmaceuticals Corporation; Pfizer Inc.; Piramal Imaging;Servier; Takeda Pharmaceutical Company; and Transition Therapeutics.The Canadian Institutes of Health Research is providing funds to support ADNI clinical sites in Canada. Private sector contributions are facilitated by the Foundation for the National Institutes of Health (www.fnih.org). The grantee organization is the Northern California Institute for Research and Education, and the study is coordinated by the Alzheimer’s Therapeutic Research Institute at the University of Southern California. ADNI data are disseminated by the  Laboratory  for  Neuro Imaging  at  the University of Southern California. 

\end{document}